# Machinic Surrogates: Human-Machine Relationships in Computational Creativity

**Ardavan Bidgoli[1], Eunsu Kang[2], Daniel Cardoso Llach[1]**
[1] Computational Design Laboratory, [2] School of Computer Science
Carnegie Mellon University
Pittsburgh, Pennsylvania, USA
{abidgoli, eunsuk, dcardoso} @andrew.cmu.edu

## Abstract

Recent advancements in artificial intelligence (AI) and its sub-branch machine learning (ML) promise machines that go beyond the boundaries of automation and behave autonomously. Applications of these machines in creative practices such as art and design entail relationships between users and machines that have been described as a form of "collaboration" or "co-creation" between computational and human agents [1, 2]. This paper uses examples from art and design to argue that this frame is incomplete as it fails to acknowledge the socio-technical nature of AI systems, and the different human agencies involved in their design, implementation, and operation. Situating applications of AI-enabled tools in creative practices in a spectrum between automation and autonomy, this paper distinguishes different kinds of human engagement elicited by systems deemed "automated" or "autonomous." Reviewing models of artistic collaboration during the late 20th century, it suggests that collaboration is at the core of these artistic practices. We build upon the growing literature of machine learning and art to look for the human agencies inscribed in works of "computational creativity", and expand the "co-creation" frame to incorporate emerging forms of human-human collaboration mediated through technical artifacts such as algorithms and data.

## Introduction

The application of AI-enabled tools in creative practices raises questions regarding the relationship between humans and their tools. Compared with traditional tools, i.e., a paintbrush, computer programs that leverage artificial intelligence models to accomplish a task, are not passive objects to facilitate artists' creative expression. These tools are designed and implemented to intervene in the course of actions and contribute to the creative process. This paper explores the relationship between human agents and AI-enabled tools and aims to outline the different roles that each one plays in a creative task.

Recent scholarship from the field of design and technology studies has shed light on how since the postwar, computers have enacted different roles in popular imaginaries of design —sometimes appearing as "perfect slaves" poised to liberate designers from labor, and sometimes as "collaborative partners" creatively contributing to the design process [3, p. 54]. Imagined as perfect slaves, machines are deprived of any creative agency. While the tool is an essential part of the process, and introduces new horizons to explore, creativity remains a distinctly human attribute. Imagined as collaborative partners, by contrast, computers are endowed with human attributes including creativity, judgment, and even sense of humor [3, p. 79]. Missing from these two frames is the recognition of these technologies' infrastructural scale and their nature as designed artifacts and thus as enactments of human intent [3, p. 149]. This recognition is crucial for an engaged critical as well as creative practice of computational art and design.

Examining artistic collaborations and cross-disciplinary case-studies in the realm of artificial intelligence and design, this paper emphasizes the human agencies involved in systems conventionally represented as autonomous. We propose the term "machinic surrogate" to highlight the intentional (even if at times unpredictable) nature of these systems. The final section discusses different forms of authorial engagement enabled by these "machinic surrogates."

## Computational Creativity

Fueled by the recent advancements in the realm of AI and ML, "Computational Creativity" (CC) studies autonomous generative systems that can produce "creative products" [4, p. 197] in domains including art, music, literature, and mathematics. [5]

Computational Creativity can be traced back to the early AI proponents who were promising human-level intelligence "embedded" in computer programs and transferring human skills into the machines. The early literature of AI is permeated with the techno-optimistic and long-awaited promise of computers capable of duplicating human expertise as well as models to "elucidate" human skill and actions. [6, 7]

Despite the autonomy label, the human agency plays a critical role in the realm of CC. These tools are the outcomes of joint efforts by an assembly of human agents, constitutes of researchers, developers, and designers who collectively crafted them. CC tools that leverage ML algorithms are heavily influenced by the process in which their training data sets have been designed and collected by human agents. Inevitably, CC tools serve as a proxy to reflect the skills, decisions, and biases of the human agents behind them. The agency of these "machinic surrogates" is derived from their "human inspirers" agency. [8]

## Co-creation with Machines

In recent years, several artists repurposed tools that have been primarily developed by ML researches and adapted them to serve in their creative practices. The application of these tools in creative practices entails a relationship in which both the human agent and the machine contribute to the decision-making process. Both contributors can impose their decisions to initiate or change the course of actions and outcomes. This is a shift from the perspective that credits the human agent as the sole author and source of creativity.

To address this type of relationship, several scholars used co-creation model. It has drawn increasing interest in recent years among the human-computer interaction community. [1, 9, 10] As a broad term, co-creation refers to any act of collective creativity among human agents in different fields, including but not limited to design, public relation, business, and product development. [11]

In the realm of CC, co-creation refers to the joint effort of the human agents and machines to engage in a creative practice. The outcomes "cannot be ascribed either to the human or to the computer alone and surpasses both contributors' original intentions." [12, p. 137] Some scholars propose this relationship as analogous to the relationship between a "visionary" and a "doer", i.e., an art director and a graphic artist, or an orchestra conductor and the players. [13] In this capacity, machine demonstrates some level of autonomy that may be perceived as a form of agency. Thus, it might be possible to consider the relationship between human agents and CC tools as a form of collaboration between a human agent and a machinic agent.

As co-creation raises more interest among the scholars of human-computer interaction (HCI), it is illuminating to compare it with another form of companionship between multiple agents in creative practices, artistic collaboration.

## Collaboration in Art

During the second half of the 20[th] century, many artists sought for new means of self-representation, breaking the traditional stereotype of the individual lonely artist waiting for inspiration to strike. Artists were questioning individual identity as an "index of the self" and started exploring new forms of identity and authorship. They found collaboration as an opportunity to manipulate the artist's identity to transform it from an individual one into a "composite subjectivity." Collaborative art attracted significant attention between the 1960s and 1970s and facilitated the transition of modern art to post-modern art in that era. [14]

Collaboration is described as "… a well-defined period of time during which two or more artists network their [mutual] interests, desire, and capacities on the basis of their shared interest in the common exploration of a topic or issue." [15, p. 94] It serves as a means to push the creative boundaries and inspiration for collaborators. From his point of view, the collaborators' complementary and unexpected contributions push the results beyond the capacities of each one. [4]

Collaboration forges a new identity beyond the sum of their individual identities and challenges their individual authorship. [16] It is not a mere "merger of two hands", but it is a more profound mutual effort that goes beyond each artist's signature style and creates a "third artistic identity superimposed over and exceeding the individual artists." [14, p. 179]

Diversity in backgrounds and identities is a critical ingredient that renders collaboration fruitful and necessary. In that sense, collaboration is a "cross-cultural dialogue". The initial diversity among the collaborators will eventually erode through the association between them and paves the ground for the creation of a new identity. [17]

This conception of collaboration can be associated with the ideas of French anthropologist, Claude Levi-Strauss. He emphasizes the importance of the "emerging identity" in collaboration. Levi-Strauss addresses the process in which a new identity emerges in collaboration: "… in the course of … collaboration, they gradually become aware of an identification in their relationships whose initial diversity was precisely what made their collaboration fruitful and necessary". [17, p. 533]

Some of the scholars who define their definition of collaboration based on Lévi-Strauss's thoughts suggest that relationships which are not aimed to form a new identity or fail to do so, are not collaboration. For example, the relationship between an artist and its craftsmen is usually not intended to form such an identity. [18] The short joint projects among artists can also be excluded from collaboration definition. These projects barely scratch the surface of individuals' "authorial signature style" and fail to shape a unique identity. [14, pp. xii-xiii]

Scholars in the history of art have observed and studied several collaboration efforts during the 1960s and 1980s that successfully formed their unique identities through long-term companionship. For example, Marina Abramovic and Ula formed a long-lasting collaboration by recreated themselves as a "third identity", or as they used to describe it the "two-headed body". [14, p. 180]

## Human-machinic surrogate collaboration

Artistic collaboration, as we described above, is not directly applicable to the relationship between human agents and cc tools. Casting such a relationship as collaboration entails a basic assumption: associating identity with these tools. It is essential to determine the origin of this identity.

We argue that this identity is not derived from the machine, nor the algorithm that drives it. It is originated from the "human inspirers" and reflects the identity of the toolmakers who contributed to its development. However, it is worth mentioning that the tool is not a perfect one-to-one mapping of human knowledge, skills, or creativity. It is a re-creation of these features, situated in the algorithm and its hardware with respect to all the limitations.

From this point of view, co-creation is a special case of collaboration where the tool acts as a "machinic surrogate" to represent the identity of its "toolmakers". In synergy with the artists' identity, this surrogacy flourishes in the form of an emergent identity.

## Machinic Surrogacy in Practice

In this section, we first introduce three recent projects in which various AI-enabled tools surrogate authors' and/or users' agency. In each project, the bespoken AI-enabled tools that have been developed and/or modified by the artists serve as more than passive tools or automated systems. Combined with their hardware apparatus, they form a proxy that let the artists or audiences collaborate with the toolmakers through a machinic surrogate.

Each project accounts for a slightly different variation of machinic surrogacy by opting for a different interaction or user engagement model. This arrangement helps us to investigate different aspects of machinic surrogacy.

**DeepCloud:** The first author in collaboration with Pedro Veloso developed a data-driven modeling system that enables users to quickly generate new objects from a given class of objects, i.e., tables, chairs, cars. Users can interact with the graphical user interface to rapidly generate new objects that did not exist in the training dataset. (Figure 1) [19]

The machine learning back-end leverages an Autoencoder (AE) which was originally developed by Achlioptas et al. [20] The AE was trained on thousands of point cloud samples from different classes of the ShapeNets dataset. [21] The novel advantage of using this machine learning model crystalizes in the non-parametric representation of point cloud objects. During the learning process, this representation is being encoded in the AE by registering patterns and finding similarities in the data sets.

Serving as a generative model, the AE can generate new instances of each class based on the user's inputs. Users can move the sliders and rotate the knobs on a physical MIDI mixer to either manipulate different features of a given object or mix multiple ones to create a new object. [19]

In DeepCloud, machinic surrogacy is exhibited in the various parameters which were set by the original toolmakers and modified by the first author and Veloso, namely the AE architecture, the data set selection, and choice of hyperparameters. While the user is free to directly interact with the apparatus and make its own decisions to shape new objects, the range of outcomes and the design space is confined by these factors.

**My artificial muse**: Mario Klingemann, Albert Barqué-Duran, and Marc Marzenit took a different approach into the relationship between the artist, algorithms, and audiences. They chained different ML models to create a "pose-to-image tool". They started by training an ML model to extract body poses from still images. Then they trained their

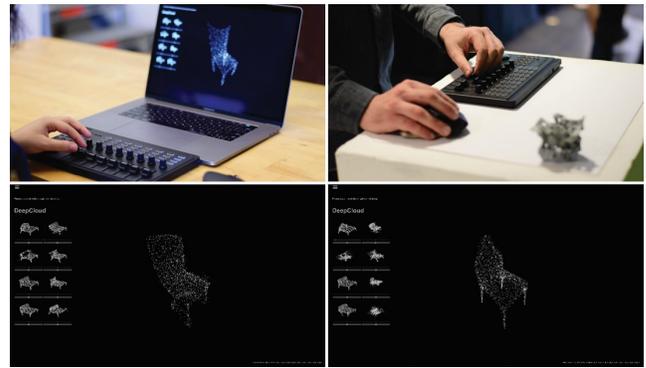

Figure 1- *DeepCloud*, the physical interface (top), the interface (bottom). Images from [19]

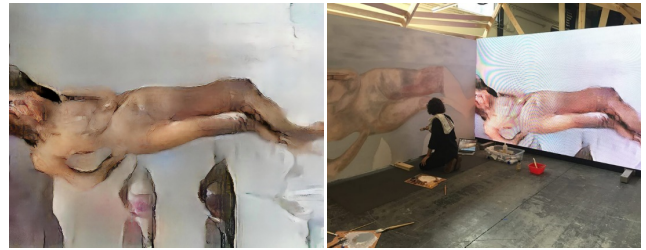

Figure 2- *My Artificial Muse*, the generated image (left), the performance (right). Images from [22, 23]

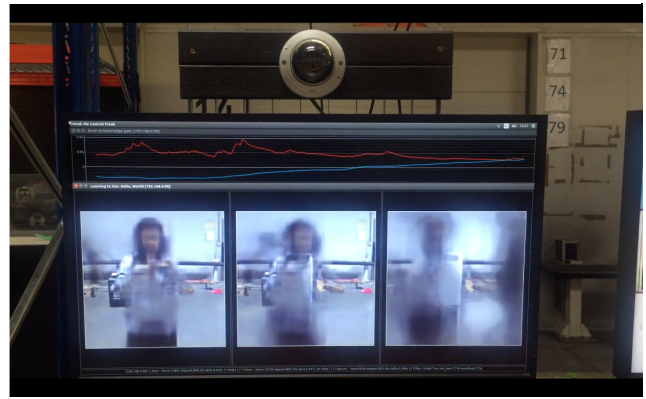

Figure 3- *Learning to see*: *Hello, World!* Image from [24]

Generative Adversarial Networks (GANs) to generate muses based on a given pose.

During a three-day performance, audiences could vote on their favorite pose to be used as the input for the generation of a digital image of an artificial muse. At last, Barqué-Duran painted the image on a 4 x 2.7m canvas. [22] (Figure 2)

Compared with *DeepCloud*, in this project, the toolmaker, performance artist, and the audiences all engage in a creative practice collectively. As Barqué-Duran states, they "collaborated" with this tool to generate the muse. [23]

While Klingemann and other toolmakers were not directly participated in the performance, their machinic surrogacy was directly impacting every aspect of it. The tool was informed by choices that they made in advance, i.e. selection of the ML models to chain, fine-tuning them, curating the training data sets, and several other factors. Although they deliberately let the audiences chose the final pose, the process of generating the muse was heavily influenced by their surrogate agency.

**Learning to See (Hello, World!)**: Memo Akten used various generative machine learning models, that have been primarily developed for ML research, to make a series of works titled *Learning to see*. [24] In *Hello, World!*, he used a Convolutional Variational Autoencoder, [25] and designed a specific model of interaction between the ML model and the audiences by providing them the opportunity to retrain the Convolutional Neural Network (CNN) in realtime. As an audience feeds the surveillance camera with visual inputs, the back-end CNN model is being trained in realtime to recognize patterns and shapes. However, the effect of each input signal gradually vanishes through the time as new audiences feed the model with new learning samples. (Figure 3) [26]

Compared with the other two projects, *Hello, World!* is best described as an interactive installation that calls for audiences' contribution as the source of training data. In the previous two projects, the AI-enabled tool was a surrogate for the absent human agents. In contrast, in *Hello, World!* the tool is pushed one step further to serve as a surrogate for the present audience. It faces the audience with its own decisions from a moment before. The toolmakers, artist, and audience are all contributing to the emergence of new characteristics that cannot be detached from their inputs.

## Humans' Role in Collaboration with Machinic Surrogates

As discussed in the previous section, human agents can approach AI-enabled tools in various capacities and adopt different roles during each phase of a project's life cycle. For example, designing an ML model to empower a CST is a creative activity that requires a wide range of technical skills. It demands for experience, creativity, improvisation, and constant decision making. Developers who design and implement such tools inject their knowledge, skill, and agency to their craft. In this sense, developers act as toolmakers and authors. A classic example is the work of Ian Goodfellow et al. who invented GANs. [27] They designed and created a novel architecture for generative models in machine learning that has been extensively used for creative applications by various artists.

On the other hand, there are artists, or practitioners who use currently available ML models and repurpose them to serve their creative practices. Due to the open-source availability of a wide range of these ML models, artists and developers can modify and adjust them for their own specific goals. In such cases, these artists and practitioners are fading the borders between the user and the original author. They inject their own agency into the tool by modifying the architecture, moderating the training datasets, fine-tuning the hyperparameters, and proposing new forms of user interactions. They demonstrate a level of authorship and agency, but it is generally confined to the boundaries of the models that are available for them. The artists that have been introduced in the previous section, in the context of the discussed projects, fit in this range.

These two groups are occupying the apex of the tool making pyramid and creating machinic surrogates, either to enhance the creative practice of their own or facilitate the others'.

Some artists choose to be the users of the machinic surrogates. They generate creative contents by using AI-enabled tools "as they are" or use off-the-shelf ones. They treat these tools as black boxes with limited knowledge of their internal mechanics. Hence, they do not simply co-create with these tools, they interface the previous two groups, the toolmakers, through a machinic surrogate. An artist who uses DeepCloud to develop an early design sketch is an example of this group. The artist might not be aware of the behind-the-scene orchestra, but it enjoys the new possibilities that the tool brings to the table.

Human agents can also evaluate and judge the outcomes to determine which instances are qualified to be considered as artwork. In many cases, the artist itself serves this role. However, professional curators or even a group of audiences, through crowdsourcing, may make such decisions.

The other role that human agents can occupy is the role of the audiences. From those who enjoyed the early outcomes of "Deep Dream" [28] to the individuals who visited the "Gradient Descent" exhibition, [29] these individuals are the audiences who enjoyed AI-assisted artworks.

Various aspect of human agents' role can be best illustrated through a recent project by the *Obvious* team. This Paris-based trio of artists uses GANs to generate images that mimic classic portraits. They state that their role is limited to moderating the inputs and selecting the best ones among the pool of outputs. They signed the portraits with the famous GANs objective function to emphasize their argument that the piece is generated by an algorithm. But a closer look reveals a chain of machinic surrogacies and various hum in action to create this piece.

As the trio stated, their machine learning algorithm was inspired by *Art-DCGAN* which was modified, implemented, and trained by a young developer, Robbie Barrat. As Barrat describes in the projects' GitHub page, it is a "modified version of Soumith Chintala's torch implementation of *DCGAN* with a focus on generating artworks." [30] Together, this chain of researchers and developers can be credited as the original author and the toolmaker.

The Obvious team trained the algorithm on a dataset of classic portraits to generate a series of portraits, playing a hybrid role of toolmaker-artist. The resulted tool serves as a surrogate for ML experts, developers, and the Obvious team.

After generating a collection of outcomes, a team of curators and experts at Christie's evaluated them to pick one to be auctioned, acting as a moderator/curator. Eventually, the outcome titled "Portrait of Edmond Belamy" presented to art enthusiasts as the audiences and was sold in an auction in October 2018.

There are no solid boundaries between the aforementioned roles. Human agents may adopt different roles,

switch between them, or adopt multiple ones simultaneously. For example, in *Learning to see*, the audiences are also collaborators by feeding the ML model with new sets of training data and influencing the model's behavior. Another example of this dynamic role shifting is demonstrated in "Sketch RNN". The training dataset of this model was constantly collecting samples from the audience who were interactions through an online game titled "Quick, Draw!". The users were not only entertained by the outcomes, but they were constantly enriching the learning dataset by their inputs. Although due to the large number of co-creators, the contribution of each individual was subtle. [31]

## Authorship and Ownership

When an artist uses an AI-enabled tool, it indirectly collaborates with a group of agents through a machinic surrogacy. Therefore, like any other form of artistic collaboration, the authorship could be associated with all the engaged agents.

There are various point of views to this topic, spanning from crediting the creativity solely to the agent who uses the tool, to acknowledging the ability of tools to create. In the early years of computer-aided design, Steven Coons believed that machines can be programmed to generate stylized content, i.e., music with the style of Vivaldi, but the "creative act" has been already performed. He argued that the machine is only extracting the "skeletal structure" that the created act was performed in it. [3] In recent years, artists like Klingemann assert that these tools are just a means in the hand of creative artists. He compares these tools with music instruments, emphasizing that the instrument is not an artist. [32] In contrast, specifically in popular media, machines have been credited as the agents which create art pieces.

The machinic surrogate perspective takes the middle ground between the two extreme ends of this spectrum. It distinguishes the AI-enabled tools from conventional tools of artistic expression, i.e., a musical instrument, due to their unique capacity to act as a surrogate for the human agent. Hence, because of the same reason, it does not solely credit the tool as the author either.

The ambiguity in ownership is another controversial source of discussion in this field. During the 1990s, the practice of relational art demonstrated that an artwork is not always an attainable asset and sometimes it cannot be confined in one's possession. [33] The same applies to the application of AI-enabled tools.

On one side, the generated artwork is the result of a probabilistic model. Thus, each edition of the work is generated unique and cannot be reproduced later. However, these unique editions can be cloned digitally in a countless number of copies. This defies the values associated with the rarity of an art piece. One approach is to treat these pieces in the same fashion as photography and printmaking, where the results are multipliable. In such cases, the rarity should be "manipulated" and "produced". [34] There are efforts to employ different technological solution, i.e., blockchain, to keep track of the authentic editions of digital art pieces and control unauthorized copies in the market.

## Conclusion

Applications of AI and ML in creative practices can be seen as a form of collaboration between human agents mediated by technical artifacts such as algorithms and data, in which groups interface with others through what we have termed in this paper "machinic surrogates." Despite the claims on "autonomous creativity" of machines, it is the agency of the authors and toolmakers which is crystallized in the tool, creative process, and the outcomes in the examples we have studied. Human agents adopt various roles in the life cycle of a co-creation scenario, from being the original author of the algorithms, to enjoy the results as an audience.

The concept of machinic surrogacy might be also applicable to future AI advancements. Algorithms that can generate algorithms are among the challenging fields for machinic surrogacy since the human agent's contribution decays in each iteration. The fact that artistic forms such as shapes, color schemes, textural details of generated results are often not predictable by the human partner presents the possibility of minimal amount of surrogacy by the future development of machine learning.

We narrowed the scope of this paper to a subset of projects that have used ML algorithms during the past few years (2010-2018). A possible next step for this paper is to extend the discussion to other branches of AI as well as autonomous computer creativity tools. Not included in this paper but we notice current efforts of developing "creative" algorithms that may result in much amplified role of the machine. We are also looking forward to studying those examples in the future.

## Acknowledgments

The authors would like to express their gratitude to CMU's Machine Learning department and the Dean's Office of School of Computer Science for its generous support. The authors would like to thank Memo Akten and also Obvious team for their feedbacks that helped us to accurately represent their projects.

## Authors Biographies

**Ardavan Bidgoli** is a Ph.D. candidate in Computational Design at the School of Architecture, Carnegie Mellon University. His research is focused on machine learning generative models and human-machine co-creation in creative practices. He is the robotics fellow at the Computational Design lab (Code Lab) and routinely contributes to the architectural robotic research at the Design Fabrication lab (dFab) where he teaches Introduction to Architectural Robotics. His research has been published and presented in ACADIA, CAADRIA, RobArch, and NeurIPS. He had been collaborating with Bentley Systems as well as Autodesk's OCTO team at Pier 9 and BUILD space facilities. Ardavan has a Bachelor of Architecture and a Master of Architecture from the University of Tehran, Iran, and a Master of Architecture in Design Computing from The Pennsylvania State University.

**Dr. Eunsu Kang** is a Korean media artist who creates interactive audiovisual installations and AI artworks. Her current research is focused on creative AI and artistic expressions generated by Machine Learning algorithms. Creating interdisciplinary projects, her signature has been the seamless integration of art disciplines and innovative techniques. Her work has been invited to numerous places around the world including Korea, Japan, China, Switzerland, Sweden, France, Germany, and the US. All ten of her solo shows, consisting of individual or collaborative projects, were invited or awarded. She has won the Korean National Grant for Arts three times. Her researches have been presented at prestigious conferences including ACM, ICMC, ISEA, and NeurIPS. Kang earned her Ph.D. in Digital Arts and Experimental Media from DXARTS at the University of Washington. She received an MA in Media Arts and Technology from UCSB and an MFA from the Ewha Womans University. She had been a tenured art professor at the University of Akron for nine years and is currently a Visiting Professor of Art and Machine Learning at the School of Computer Science, Carnegie Mellon University.

**Dr. Daniel Cardoso Llach** is Associate Professor at Carnegie Mellon University, where he teaches architecture, directs the Master of Science in Computational Design, and co-directs the Code Lab, a multidisciplinary laboratory focusing on critically exploring design technologies. He is the author of *Builders of the Vision: Software and the Imagination of Design* (Routledge), which identifies and documents the theories of design emerging from postwar technology projects at MIT, and traces critically their architectural repercussions. He is a Graham Foundation grantee and the curator of a recent exhibition on the history and possible futures of computational design. His writings have been published in journals including *Design Issues*, *Architectural Research Quarterly* (ARQ), and *Thresholds*, among others, and in several edited collections, including *The Active Image: Architecture and Engineering in the Age of Modeling* (Springer 2017) and the forthcoming *DigitalSTS: A Handbook and a Fieldguide* (Princeton 2019). Daniel routinely lectures and teaches workshops around the world. He holds a Bachelor of Architecture from Universidad de los Andes, Bogotá, and a Ph.D. and MS (with honors) in Design and Computation from MIT. He has also been a research fellow at Leuphana (MECS), Germany, and a visiting scholar at the University of Cambridge, UK.